\documentclass{article}

\usepackage{PRIMEarxiv}

\usepackage[utf8]{inputenc} % allow utf-8 input
\usepackage[T1]{fontenc}    % use 8-bit T1 fonts
\usepackage{hyperref}       % hyperlinks
\usepackage{url}            % simple URL typesetting
\usepackage{booktabs}       % professional-quality tables
\usepackage{amsfonts}       % blackboard math symbols
\usepackage{nicefrac}       % compact symbols for 1/2, etc.
\usepackage{microtype}      % microtypography
\usepackage{lipsum}
\usepackage{fancyhdr}       % header
\usepackage{graphicx}       % graphics
\graphicspath{{media/}}     % organize your images and other figures under media/ folder
\usepackage{cite}
\usepackage{bm}
\usepackage{amsmath,amssymb,amsopn,mathtools}
\title{Efficient machine-learning surrogates for large-scale geological carbon and energy storage
}

\author{
  Teeratorn Kadeethum, Stephen J. Verzi, Hongkyu Yoon \\
  Sandia National Laboratories, New Mexico, USA \\
}

\begin{document}
\maketitle

\begin{abstract}
Geological carbon and energy storage play a pivotal role in achieving a net-zero carbon emission for mitigating climate change impact. Nevertheless, these storage endeavors confront uncertainties tied to geological formations and operational constraints, which may lead to induced seismic events or groundwater contamination. To tackle these challenges, machine learning (ML) algorithms emerge as a promising solution for optimizing operations and preventing such complications.

Despite their potential to enhance our understanding of geological carbon storage, ML-driven approaches for subsurface physics encounter a substantial hurdle: the considerable computational resources needed for large-scale industrial analyses and management of acceptable accuracy. In this context, we propose a specialized ML model tailored for geological storage capable of efficiently managing extensive reservoir models without imposing overwhelming computational demands.

While autoencoder-based or neural-operator (NO)-based models have exhibited promise in subsurface physics, their training typically demands excessive computational resources, especially with large-scale applications. We have developed a method to reduce the computational cost of training deep NO models for carbon storage to address this issue. This method separates the training and prediction domains, meaning that the sizes of these domains are different. This method uses domain decomposition techniques to utilize a subset of spatio-temporal points during the training phase. It employs a topology embedder that links corresponding temporal and spatial coordinates in input and output spaces. This approach allows us to operate a trained model (or models) for making predictions at any points and times within the model's domain, even if they were not part of the training data (interpolation, not extrapolation). By employing this technique, we can train the models on a fraction of the computational area (with variations for each back-propagation) while still enabling them to predict the full-size field during actual operation accurately. This technique significantly enhances the efficiency of the ML model and renders it suitable for large-scale geological storage applications.
\end{abstract}

% keywords can be removed

\section{Background and motivation.}

To achieve a net-zero carbon emissions target, sustainable energy systems within the subsurface, encompassing carbon and energy storage, are imperative \cite{bickle2009geological,march2018assessment}. Achieving longevity and efficient operations in these subsurface energy endeavors necessitates a deeper comprehension of the underlying subsurface physics, the accurate estimation of subsurface properties, and a thorough grasp of their influence on subsurface systems \cite{yoon2012highly, chang20183}. The main challenge arises from the intricate nature of subsurface structures, composed of highly heterogeneous structures (e.g., porosity or permeability). This complexity presents formidable barriers to precise predictions of reactive flow, transport, and deformation \cite{yoon2012pore, matthai2009upscaling, na2017effects, kadeethum2019investigation}.

Furthermore, subsurface systems involve interrelated processes, including chemical reactions, mechanical interactions, hydrological phenomena, and thermal effects. The interaction among these coupled processes results in even more intricate and non-linear behaviors within subsurface systems \cite{nield2006convection, chang2021mitigating, kadeethum2021nonTH}. Among these complex subsurface physics challenges, multiphase flow problems stand out, particularly when dealing with complex geometries and boundary conditions. To address these challenges, one widely adopted approach is the full order model (FOM), which employs numerical techniques such as finite difference, finite volume, or finite element methods \cite{evans2012numerical, jackson:15, keilegavlen2019porepy}. However, these FOMs have significant computational demands despite their flexibility and robustness. This computational burden can constrain their applicability in real-time reservoir management, robust uncertainty quantification, or high-dimensional inverse modeling \cite{hesthaven2016certified, pasetto2013reduced}.

In this research, our objective is to enhance computational efficiency while maintaining an acceptable level of accuracy by utilizing a machine learning-based reduced order model (ROM). The parametrized ROM functions as a surrogate model defined by a set of parameters ($\bm{\mu}$) corresponding to physical properties, geometric attributes, and/or boundary conditions. This surrogate model approximates relevant quantities, such as fluid pressure or fluid saturation \cite{copeland2021reduced, choi2021space, carlberg2018conservative, kadeethum2021framework, kadeethum2021non}. In recent years, significant progress has been made in the development of accurate and efficient ROMs for subsurface energy storage, contributing to the acceleration of the energy transition \cite{shen2018transdisciplinary, wen2023real, chen2023capacity}. For instance, Kadeethum et al. (2021) \cite{kadeethum2021non} demonstrated the application of linear compression ROMs in modeling coupled hydro-mechanical processes within highly heterogeneous porous media. However, the limitations of linear subspace ROMs, based on techniques like Proper Orthogonal Decomposition (POD), became evident when addressing highly nonlinear problems, as emphasized by Kadeethum et al. (2022) \cite{kadeethum2021nonTH}. To address this challenge, researchers have turned to nonlinear compression techniques, including deep learning architectures based on autoencoders and nonlinear manifold methods \cite{fresca2021comprehensive, kadeethum2021nonTH, kim2020efficient}. These approaches have proven successful in capturing the intricacies of nonlinear phenomena.

Nevertheless, machine learning-based models still face several challenges that must be addressed. One notable issue is the bottleneck associated with data acquisition. Obtaining high-quality datasets that accurately represent the problem at hand can be time-consuming and resource-intensive \cite{obermeyer2016predicting}. Creating such datasets in specialized domains like subsurface physics or medical imaging often requires significant manual labor and expertise \cite{greenspan2016guest}. Furthermore, the computational demands of deep learning models are substantial, often necessitating access to powerful hardware resources such as Graphics Processing Units (GPUs) or Tensor Processing Units (TPUs) for both the training and deployment phases. This computational limitation can hinder the effective handling of industrial-scale reservoir models, particularly those characterized by high levels of heterogeneity, including features like faults and fractures, as well as numerous state variables like pressure and $\mathrm{CO_2}$ saturation \cite{kadeethum2022fomassistrom}.

Recently, there have been advancements in data-driven learning of solution operators for PDEs, as exemplified by DeepONet \cite{lu2021learning} and neural operator (NO) \cite{li2020neural}. These methods have demonstrated the capability to approximate PDE solutions in a forward solution context. However, it's essential to acknowledge a bottleneck associated with NOs, where each degree of freedom corresponds to discretizing the problem domain. This becomes particularly challenging as the domain size increases, as is often the case in industrial-scale reservoir models, making training impractical.

To address this issue, we propose a novel NO model that divides the computational domain into smaller subdomains during each propagation step, effectively circumventing the computational bottleneck. This approach liberates the model from being constrained by the domain size and has the added benefit of increasing the training dataset, thereby reducing data generation costs. It's worth emphasizing that this framework offers the flexibility needed to efficiently train models, even for large-scale geological carbon storage simulations.

\section{Method and model description.}

This section introduces our framework, which builds upon the DeepONet model \cite{lu2021learning,goswami2022deep}. The original DeepONet comprises two primary components: the trunk and branch networks. The trunk network handles the topological space, encompassing temporal and spatial coordinates. Conversely, throughout this manuscript, the branch network focuses on the parameter space, denoted as $\bm{\mu}$.

\begin{figure}[!ht]
   \centering
    \includegraphics[keepaspectratio, height=6.5cm]{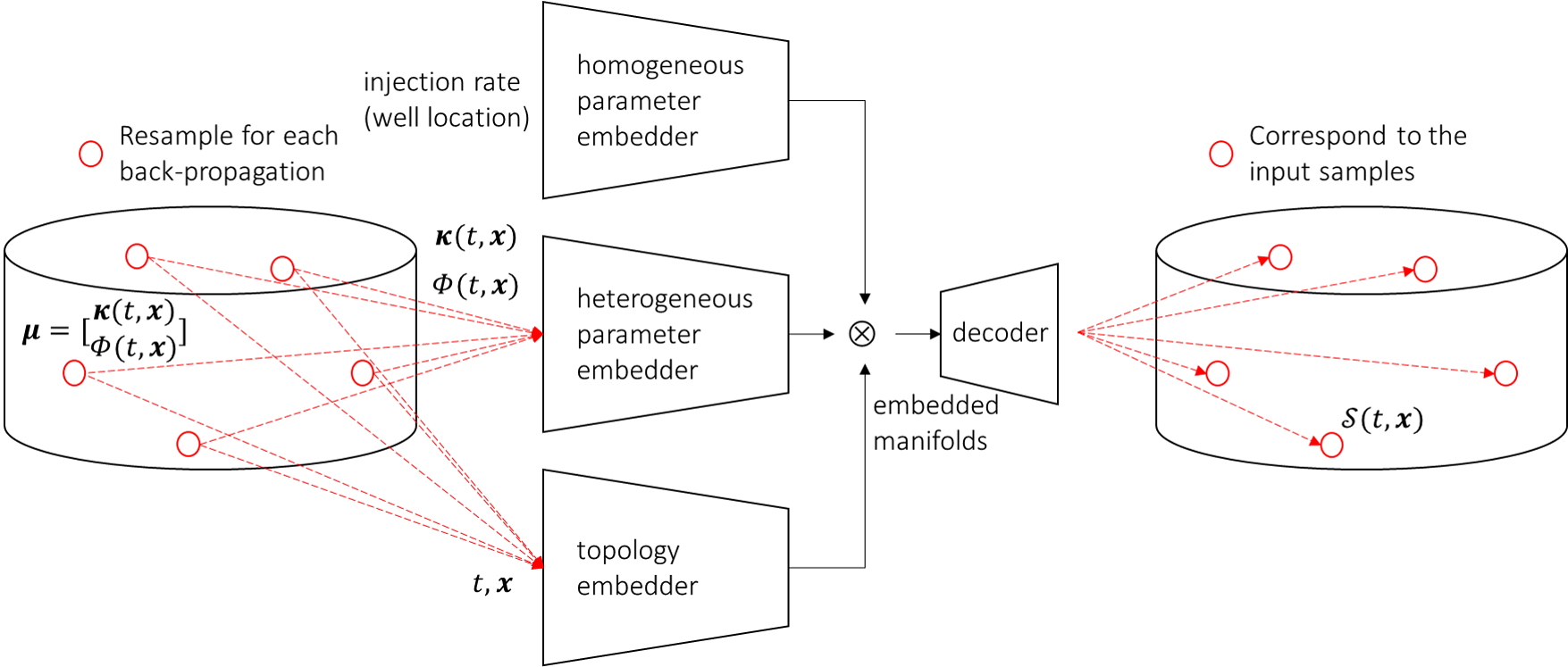}
   \caption{Schematic of the proposed improved neural operator.}
   \label{fig:model}
\end{figure}

Our model comprises three essential components, as illustrated in Figure \ref{fig:model}: a topology embedder (TE), a heterogeneous parameter embedder (HePE), and a homogeneous parameter embedder (HoPE). The TE operates similarly to the trunk network, overseeing the topological space. Its primary role involves processing temporal and spatial coordinates and compressing them into reduced or embedded manifolds.

The HePE is specifically tailored to manage high-dimensional inputs, such as porosity or permeability fields, effectively converting them into embedded manifolds. Conversely, the HoPE handles low-dimensional inputs, like injection rates or well locations, and reduces them to embedded manifolds. Once these embedded manifolds are constructed, we proceed to perform element-wise summation, and subsequently, we pass the resulting combined embedded manifold to a decoder for the reconstruction of desired state variables.

As for the model architecture, we use

\begin{align*}
\text{HoPE dimension} &= [2, 32, 32, 32, p], \\
\text{HePE dimension} &= [4, 512, 512, 512, 512, p], \\
\text{TE dimension} &= [4, 512, 512, 512, 512, p],
\end{align*}

\noindent
where $p$ is an embedded dimension, set at 250, and each number within $[\cdot]$ represents the in- and out-feature of each linear layer. To elaborate, for the first linear layer of HePE, the in- and out-feature are 4 and 512, respectively. Each linear layer is followed by the LeakyReLU activation function with a negative slope of 0.2 (except the last layer, which is subjected to the tanh activation function). Each linear layer (except the last layer) is subjected to a 30\% dropout. The decoder is a single linear layer with an in-feature of $p$ and an out-feature of 1.

% Each linear layer (except the last layer) is followed by layer normalization \cite{ba2016layer}. Layer normalization is a technique used in deep neural networks to address the challenges of internal covariate shifts and improve training stability. It normalizes the inputs for a specific feature across all the data points in a single example. 

As previously mentioned, addressing the bottleneck issue in DeepONet or NOs is crucial, where each degree of freedom represents a discretization of the problem domain. This concern becomes more pronounced when dealing with larger domains, as often encountered in industrial-scale carbon storage models, making training or inferring impractical. To tackle this challenge, we have implemented a solution by subsampling the computational domain, as illustrated in Figure \ref{fig:model} - training. This involves randomly selecting a portion of the domain during each backpropagation pass, as depicted by the red circles in Figure \ref{fig:model}.

This subsampling mechanism empowers us to decode the reduced manifolds and make predictions for desired state variables at any spatial and temporal coordinate. Consequently, this approach decouples model operations from the size of the training domain, granting the model the flexibility to be trained effectively with industrial-scale reservoir models. We note that the model presented in this study relies on a sampling scheme with uniform distribution as prior. Our full manuscript will comprehensively compare different sampling schemes (e.g., Latin hypercube, Sobol, or Halton sequences). We use a subsample size of 4096 throughout this study.

For the loss function, we utilize the mean squared error (MSE)

\begin{equation}
    \text{MSE} = \| \mathbf{S} - \hat{\mathbf{S}}  \|_2
\end{equation}

as our metric. Here, $\|\cdot\|_2$ is $\mathcal{L}_2$ norm, $\mathbf{S}$ is our ground truth, and $\hat{\mathbf{S}}$ is our prediction. The training protocols closely resemble those outlined in our earlier studies \cite{kadeethum2021framework,kadeethum2022reduced}. To optimize the learnable parameters within all embedders and the decoder, we employ the adaptive moment estimation (ADAM) algorithm \cite{kingma2014adam}. The learning rate ($\eta$) is computed using the method proposed in \cite{loshchilov2016sgdr} 

\begin{equation}\label{eq:learning_rate}
\begin{split}
\eta_{c} =  &  \eta_{\min } \\ & +\frac{1}{2}\left(\eta_{\max }-\eta_{\min }\right)  \left(1+\cos \left(\frac{\mathrm{step_c}}{\mathrm{step_f}} \pi\right)\right).
\end{split}
\end{equation}

\noindent
In this context, $\eta_{c}$ represents the learning rate at step $\mathrm{step_c}$. The minimum learning rate, denoted as $\eta_{\min}$, is set to $1 \times 10^{-16}$, while the maximum or initial learning rate, $\eta_{\max}$, is chosen as $1 \times 10^{-4}$ for the outer loop and $1 \times 10^{-5}$ for the inner loop. Here, $\mathrm{step_c}$ refers to the current step, and $\mathrm{step_f}$ is the final step in the process. We note that all of these choices of the model's architecture and hyper-parameters are selected based on a comprehensive sensitivity analysis provided in our full manuscript.  

\section{Numerical examples.}

In this section, we utilize data from the Illinois Basin – Decatur Project (IBDP) in Decatur, Illinois, as detailed in the work by Finley et al. (2014) \cite{finley2014overview}, to exemplify our practical capabilities in a real-world context. The IBDP injected millions of metric tons of $\mathrm{CO_2}$, captured from biofuel production, into a deep saline subsurface over three years from November 2011 to November 2014. Post-monitoring activities have extended up to 2021. Concurrently, a commercial company, Archer Daniels Midland, has continued conducting another $\mathrm{CO_2}$ injection project since 2017. This ongoing endeavor demonstrates the commercial feasibility of $\mathrm{CO_2}$ capture and storage technology, with the primary objective of reducing greenhouse gas emissions from industrial sources.

\begin{figure}[!ht]
   \centering
    \includegraphics[keepaspectratio, height=10.75cm]{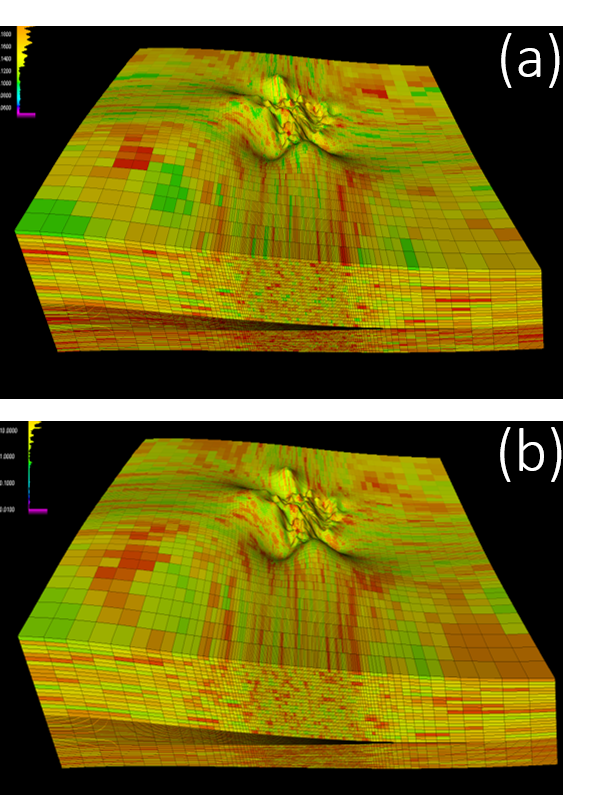}
   \caption{Illustration of input parameters: (a) permeability and (b) porosity.}
   \label{fig:siam_input}
\end{figure}

We have two input parameters in this context: a timestamp denoted as $t$ and spatial coordinates represented as $\bm{x} = [x, y, z]$. These parameters serve as inputs to the TE model. Additionally, we deal with two heterogeneous parameters, namely porosity, $\phi\left(t, \bm{x}\right)$, as illustrated in Figure \ref{fig:siam_input}b, and permeability, $\bm{\kappa}\left(t, \bm{x}\right)$, as illustrated in Figure \ref{fig:siam_input}a, which function as inputs for the HePE. Another critical input parameter is the injection rate, which the HoPE uses. \par

Our ultimate output is a state variable, denoted as $\mathbf{S}\left(t, \bm{x} \right)$, which can either represent fluid pressure ($p\left(t, \bm{x} \right)$) or $\mathrm{CO_2}$ saturation ($s\left(t, \bm{x} \right)$). We must highlight that we construct each state variable using a dedicated model. Although it's possible to consider a combined model, we have not yet explored this approach.

We possess a total of 100 samples at our disposal. For the purpose of training and testing, we have allocated 90 cases for training and 10 cases for testing. Due to the limited availability of samples, each of which demands substantial computational resources to process (days on a high-performance computing machine), we have opted to forego the utilization of a validation set in this research endeavor.

The structure of each sample is defined by a 126 $\times$ 125 $\times$ 110 grid with spatial coordinates undergoing grid refinement, along with 50 timestamps that pertain to the temporal domain. This configuration signifies that each case comprises a staggering 1.73 million spatial degrees of freedom (DOFs) $\times$ 50 temporal DOFs. 

We will showcase our findings concerning fluid pressure and $\mathrm{CO_2}$ saturation in distinct sections. To assess our results, we employ three metrics. The first metric is the root-mean-square-error (RMSE), explained in detail below

\begin{equation}
\text{RMSE} = \sqrt{\frac{1}{n_t}\frac{1}{n} \sum_{i=1}^{n_t} \sum_{j=1}^{n} (\mathrm{S}_{i, j} - \hat{\mathrm{S}}_{i, j})^2}.
\end{equation}

\noindent
Here, where \(n\) represents the number of evaluated degrees of freedom (DOFs). As previously mentioned, our framework is designed to accommodate an arbitrary number of evaluated DOFs. In this context, the total DOFs for each timestamp amount to 1.73 million to provide further clarification. We utilize only 4096 DOFs during model training, yet our model can be evaluated using any number of DOFs ranging from 1 to 1.73 million. \(n_t\) signifies the number of evaluated timestamps. It is worth noting that our model is not constrained to the training timestamps of 50; it can accommodate any number of requested timestamps. \(\mathrm{S}_{i, j}\) denotes a state variable at a specific timestamp and spatial coordinates, while \(\hat{\mathrm{S}}_{i, j}\) represents a predicted state variable at the same timestamp and spatial coordinates.

The second metric is mean-absolute-error (MAE)

\begin{equation}
\text{MAE} = \frac{1}{n_t}\frac{1}{n}  \sum_{i=1}^{n_t} \sum_{j=1}^{n} |\mathrm{S}_{i, j} - \hat{\mathrm{S}}_{i, j}|,
\end{equation}

\begin{equation}
\text{Max MAE} = \text{Max} \left( \frac{1}{n} \sum_{j=1}^{n} |\mathrm{S}_{j} - \hat{\mathrm{S}}_{j}| \right).
\end{equation}

\noindent
We note that we only evaluate Max MAE for each timestamp. The last metric is point-wise difference or difference

\begin{equation}
\text{difference} = | \mathbf{S} - \hat{\mathbf{S}} |.
\end{equation}

\noindent
It's important to mention that we exclusively compute the Difference metric for each timestamp. While we can evaluate our spatial coordinates at any desired number (ranging from 1 DOF to 1.73 million DOFs), in this context, we have opted to assess our model using the full set of 1.73 million DOFs (representing the entire domain) for direct comparison with our ground truth.

\subsection{Fluid pressure.}

\begin{figure}[!ht]
   \centering
    \includegraphics[keepaspectratio, height=12.75cm]{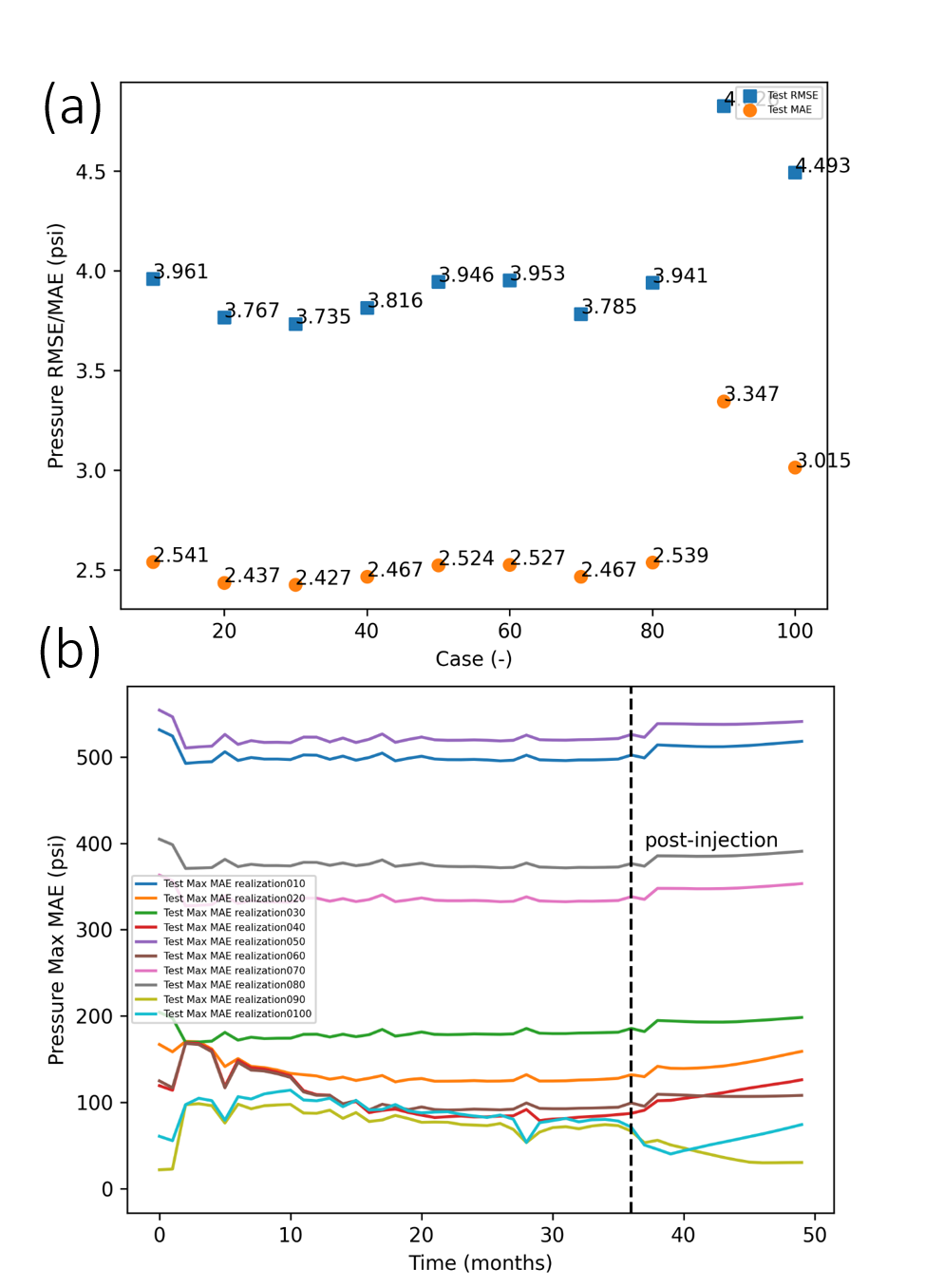}
   \caption{Pressure ($p$): (a) RMSE/MAE results of 10 test cases and (b) max MAE results as a function of time of 10 test cases.}
   \label{fig:siam_pres_results}
\end{figure}

We display the RMSE/MAE outcomes for 10 test cases of the state variable \(p\) in Figure \ref{fig:siam_pres_results}. In Figure \ref{fig:siam_pres_results}a, we observe the lowest/highest RMSE values of 3.7/4.6 psi and the lowest/highest MAE values of 2.4/3.3 psi. To contextualize these results, the average \(p\) is 3150 psi, resulting in our model yielding a relative error of 0.2 \%. Figure \ref{fig:siam_pres_results}b illustrates the maximum MAE as a function of time, with values ranging from 10 to 550 psi. Typically, the higher end of this range corresponds to well locations where the \(p\) gradient is most pronounced. \par

\begin{figure*}[!ht]
   \centering
    \includegraphics[keepaspectratio, height=10.5cm]{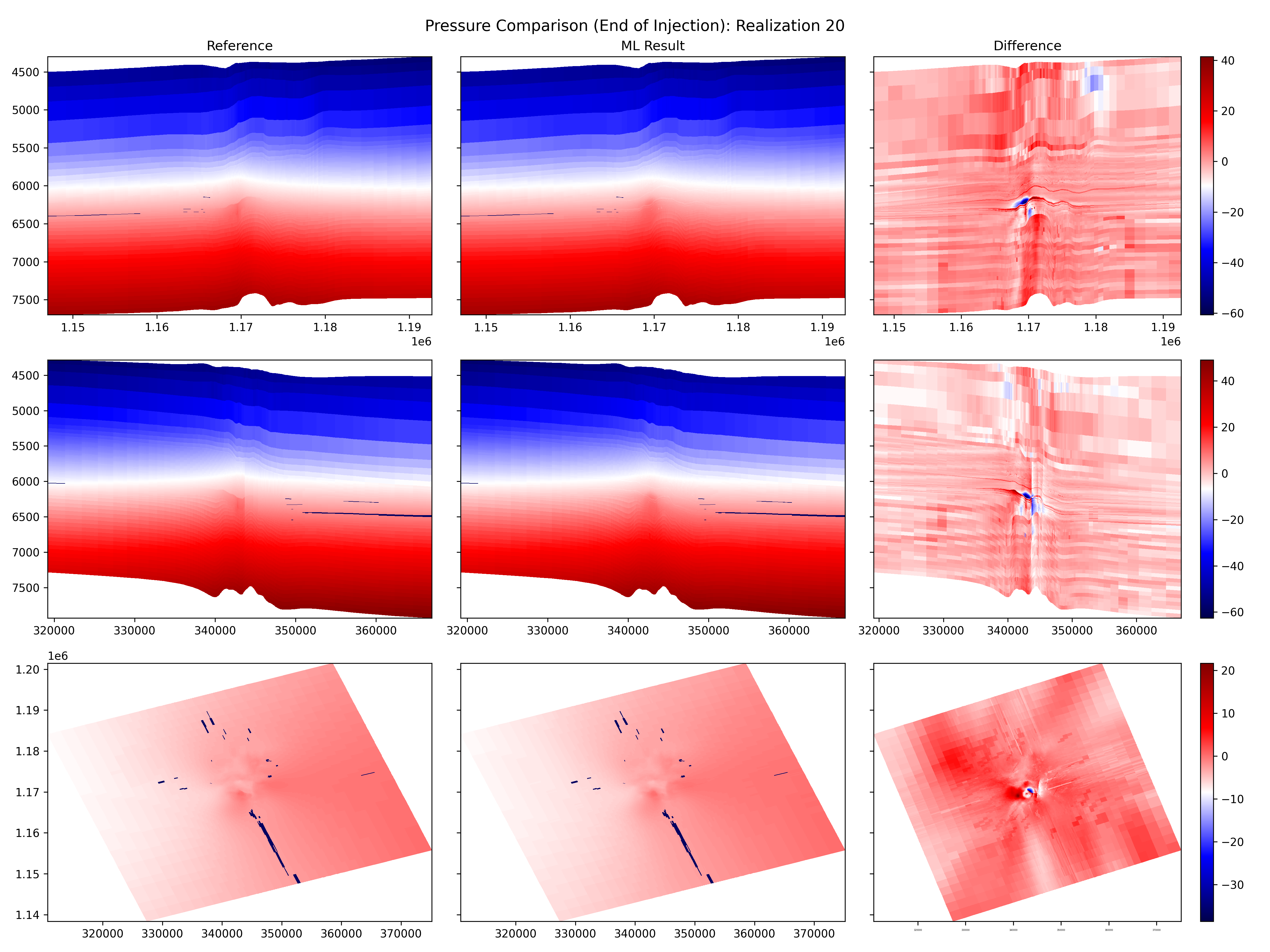}
   \caption{Pressure: point-wise comparison at the end of injection period of realization 20.}
   \label{fig:siam_pres_case2_results}
\end{figure*}

Additionally, we showcase prediction samples after the injection period for realization 20 in Figure \ref{fig:siam_pres_case2_results}. This depiction highlights that errors are most prominent in proximity to the well locations where the \(p\) gradient is particularly steep. It's worth emphasizing that this observed behavior is less than desirable, and active efforts are underway to address and mitigate this issue.  \par

\subsection{$\mathrm{CO_2}$ saturation.}

\begin{figure}[!ht]
   \centering
    \includegraphics[keepaspectratio, height=12.75cm]{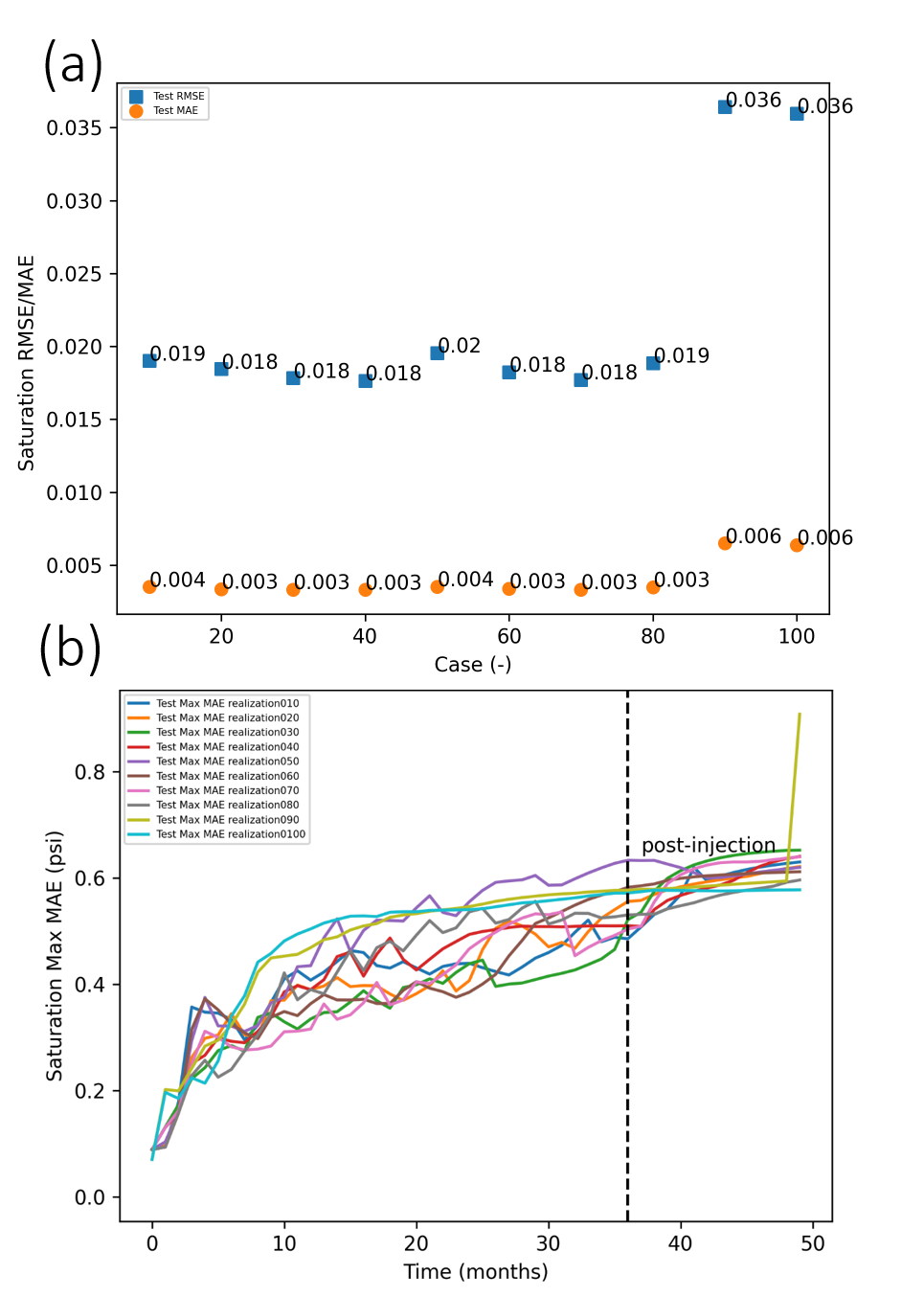}
   \caption{Saturation: (a) RMSE/MAE results of 10 test cases and (b) max MAE results as a function of time of 10 test cases.}
   \label{fig:siam_sat_results}
\end{figure}

We present the RMSE/MAE results for 10 test cases of the state variable \(s\) in Figure \ref{fig:siam_sat_results}. In Figure \ref{fig:siam_sat_results}a, we observe the lowest/highest RMSE values of 0.02/0.04 (in fractions) and the lowest/highest MAE values of 0.003/0.006 (in fractions) for \(s\). To provide context for these findings, it's important to note that \(s\) varies within the range of 0 to 1, and our RMSE/MAE values are considerably smaller than the maximum possible value.

Figure \ref{fig:siam_sat_results}b illustrates the maximum MAE as a function of time, with values ranging from 0.1 to 0.8 (in fractions). Notably, this behavior differs from that of \(p\). The higher values in this range often coincide with the $\mathrm{CO_2}$ front, where the \(s\) gradient is most pronounced. Furthermore, it's worth mentioning that while the Max MAE of \(p\) remains relatively stable over time, the Max MAE of \(s\) increases as time progresses, suggesting that errors in \(s\) accumulate with time. 

\begin{figure}[!ht]
   \centering
    \includegraphics[keepaspectratio, height=10.5cm]{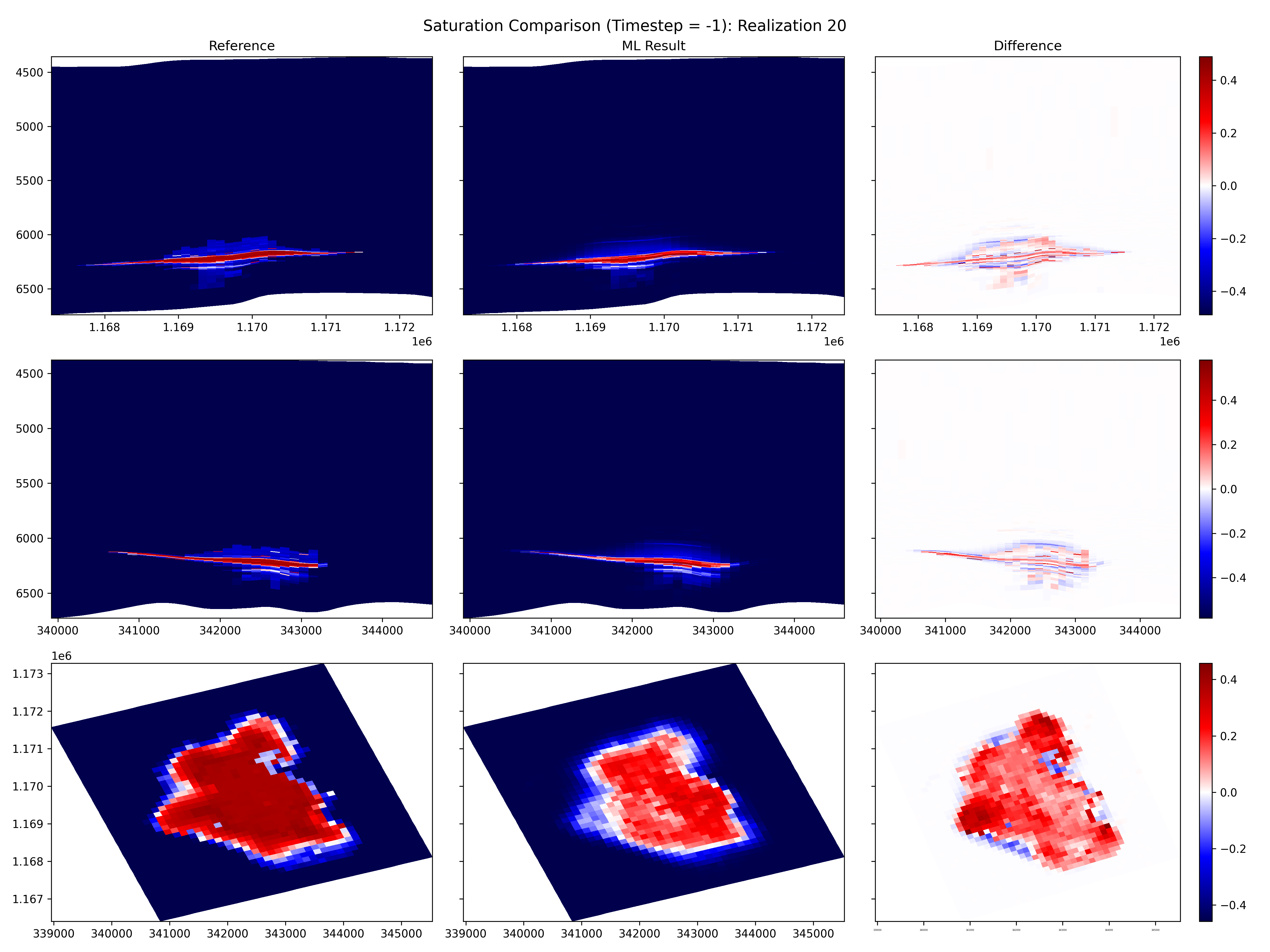}
   \caption{Saturation: point-wise comparison at the end of injection period of realization 20.}
   \label{fig:siam_sat_case2_results}
\end{figure}

Moreover, we present prediction samples post-injection period for realization 20 in Figure \ref{fig:siam_sat_case2_results}. This representation underscores the fact that errors are most conspicuous near the flood front, where the \(s\) gradient is notably steep. It is essential to underscore that this observed behavior, with errors increasing as the gradient intensifies, is less than desirable. Consequently, ongoing initiatives are in progress to tackle and alleviate this issue.

\subsection{Model complexity.}

All models are trained using a single Quadro RTX 8000. The average training time is 2700 minutes. The majority of this is allocated to subsampling operations. We note that we can also perform the subsampling operation beforehand. However, as we aim to develop an intelligent or on-the-fly subsampling technique, which we believe will improve the model accuracy with less training time, we want to report this time as is for future comparison. We note that without GPU, our framework's training time is impractical. 

The inference time required for each inquiry, specifically for one timestamp containing 1.73 million DOFs, is approximately 0.003 seconds when utilizing a single Quadro RTX 8000. Notably, one can employ a CPU for this inference operation, albeit with a slightly longer wall time. This flexibility represents one of the primary advantages of this model, as it enables inference on edge devices.

Regarding computational complexity, the model entails 1.86 million multiply-accumulate operations (MMACs) with 1.86 million trainable parameters and a compact model size of 21 MB. To provide a comparative context, the computational complexity of a model like ResNet50 stands at 4.14 billion MACs (GMACs) with a larger model size of 100 MB.

\section{Significance, impact, and future development.}

We introduce an enhanced NO model that strategically divides the computational domain into smaller subdomains during each propagation step to mitigate the computational bottleneck. This innovative approach effectively decouples model operations from the size of the training domain, offering a solution to computational challenges while simultaneously expanding the training dataset and reducing data generation costs.

As demonstrated, this methodology enables us to create a representation of the Illinois Basin – Decatur Project, an industrial-scale carbon storage initiative, using just 90 training samples, yet achieving a relative error of less than one percent on the testing set. However, our model's accuracy around high-gradient areas (around well for fluid pressure and flood front for fluid saturation) is still unacceptable. 

To further enhance the accuracy and robustness of our framework, we propose exploring various techniques such as adaptive sampling \cite{paul2015adaptive}, the incorporation of physical information \cite{kadeethum2020physics}, transfer learning through physics-guided models \cite{oommen2022learning}, or domain decomposition \cite{mcbane2021component} and subsequently integrating these methodologies into our framework.

From a perspective centered on computational costs, our next objective involves enhancing the computational efficiency of our model even further. This enhancement is pivotal to ensuring the model's suitability for both the training phase and deployment on devices characterized by limited computational resources, commonly referred to as edge devices. Our primary emphasis will continue to revolve around software optimization strategies.

One notable technique in our optimization arsenal is quantization. Quantization involves the reduction of the precision of model weights and activations, typically transitioning from 32-bit floating-point numbers to lower bit-width integers, often 8-bit, as detailed in \cite{han2015deep}. This precision reduction substantially curtails memory requirements and accelerates computational processes. Importantly, post-training quantization methods enable us to modify pre-trained models while preserving a substantial portion of their accuracy.

Efficient model size stands as a critical factor when targeting resource-constrained devices. To address this, we propose the integration of pruning methods into our framework, as elucidated in \cite{zhu2017prune}. These pruning techniques systematically eliminate redundant connections or neurons from the neural network, thereby reducing the model's dimensions while upholding its overall performance.

\section*{Acknowledgement.}
TK, SV, and HY were supported by the US Department of Energy Office of Fossil Energy and Carbon Management,the Science-informed Machine Learning for Accelerating Real-Time Decisions in Subsurface Applications (SMART) Initiative. Sandia National Laboratories is a multi-mission laboratory managed and operated by National Technology \& Engineering Solutions of Sandia, LLC (NTESS), a wholly owned subsidiary of Honeywell International Inc., for the U.S. Department of Energy’s National Nuclear Security Administration (DOE/NNSA) under contract DE-NA0003525. This written work is authored by an employee of NTESS. The employee, not NTESS, owns the right, title and interest in and to the written work and is responsible for its contents. Any subjective views or opinions that might be expressed in the written work do not necessarily represent the views of the U.S. Government. The publisher acknowledges that the U.S. Government retains a non-exclusive, paid-up, irrevocable, world-wide license to publish or reproduce the published form of this written work or allow others to do so, for U.S. Government purposes. The DOE will provide public access to results of federally sponsored research in accordance with the DOE Public Access Plan.

%Bibliography
\bibliographystyle{unsrt}  
\bibliography{references}

\end{document}